\documentclass[a4paper]{ISOStyle}
\usepackage{epsfig}
\usepackage{graphicx,subfigure}

\def\eps@scaling{.95}
\def\epsscale#1{\gdef\eps@scaling{#1}}

\def\plotone#1{\centering \leavevmode
    \epsfxsize=\eps@scaling\columnwidth \epsfbox{#1}}

\def\plotfiddle#1#2#3#4#5#6#7{\centering \leavevmode
    \vbox to#2{\rule{0pt}{#2}}
    \includegraphics{#1}}

\begin{document}

\title{FAR-INFRARED EMISSION FROM DUST IN NORMAL GALAXIES}

\author{Richard J. Tuffs\inst{1} \and Cristina C. Popescu\inst{2,3,4}}
  \institute{
     Astrophysics Division, Max-Planck-Institut f\"ur Kernphysik,
     Saupfercheckweg 1, 69117 Heidelberg, Germany 
     Richard.Tuffs@mpi-hd.mpg.de
  \and
     The Observatories of the Carnegie Institution of Washington, 
     813 Santa Barbara Str., Pasadena, 91101 California, USA
     popescu@ociw.edu
  \and
     IPAC(Caltech/JPL), 770 S. Wilson Avenue, Pasadena, California 91125, USA
  \and Research Associate, The Astronomical Institute of the 
       Romanian Academy, Str. Cu\c titul de Argint 5, Bucharest, Romania
}

\maketitle 

\begin{abstract}

We review the morphological and spectral energy distribution 
characteristics of the dust continuum emission (emitted in the 40-200$\,{\mu}$m
spectral range) from normal galaxies, as revealed by detailed ISOPHOT mapping 
observations of nearby spirals and by 
ISOPHOT observations of the integrated emissions from representative 
statistical samples in the local universe. 


\end{abstract}

\section{INTRODUCTION}

The sensitivity of ISO and its spectral grasp extending 
to 200$\,{\mu}$m made it the first observatory capable of 
routinely measuring the infrared emission corresponding to 
the bulk of starlight absorbed by interstellar dust 
in ``normal''\footnote{We use the term ``normal'' to denote star-forming 
systems not undergoing a starburst, and not dominated by AGN 
activity.} galaxies. Here we review ISO's view of the 
morphological and spectral energy distribution (SED) characteristics of the 
dust continuum emission (emitted in the 40-200$\,{\mu}$m spectral range) 
from normal galaxies, and its interpretation.
In this review we only discuss the results from the ISOPHOT instrument (Lemke
et al. 1996) on board ISO.
For the spectral observations of these systems we refer
to the review at this meeting by Helou. 

\begin{figure}[htb]
\includegraphics[scale=0.5]{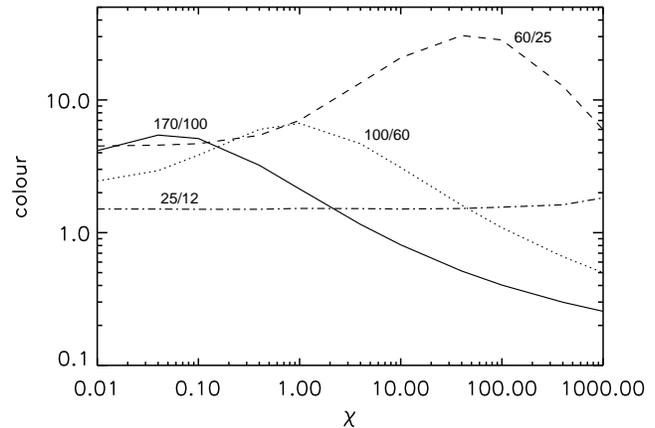}
\caption{Predicted colour ratios for standard filter combinations 
170/100$\,{\mu}$m (solid line); 100/60$\,{\mu}$m (dotted line);
60/25$\,{\mu}$m (dashed line) and 25/12$\,{\mu}$m (dot-dashed line)
as a function of the strength of the local ISRF $\chi$, where 
$\chi=1$ near the sun. The calculations were made for 
spherical grains of astrophysical silicate, with the optical properties
given by Laor \& Draine (1993). A grain size distribution
$n(a)\,da\,\propto\,a^{-3.5}$ was assumed, where $a$ is the grain radius 
($0.001\,\le\,a\,\le\,0.25\,{\mu}$m). 
The colour of the radiation field illuminating 
the grains is fixed to that determined for the solar neighbourhood
by Mezger, Mathis \& Panagia (1982).
}
\end{figure}

Although technically more demanding than observations
in the Mid-Infrared (MIR) regime, only observations in the 
Far-Infrared (FIR) directly probe the role played by dust in the
energy budget of star-forming galaxies. 
All star-forming galaxies are at least 
in part optically thick in the ultraviolet (UV)-optical regime, and the 
absorbed 
energy is predominantly re-radiated in the FIR. 
But the real investigative power of FIR astronomy
lies in the fact that even for optically thin components of the 
interstellar medium, the large grains which dominate the FIR 
emission are in (or near to) equilibrium with the ambient interstellar
radiation field (ISRF). Therefore, the grains act as test 
particles with FIR colours characteristic of the intensity and colour
of the ISRF. 
This is illustrated in Fig.~1, which shows the predicted variation
of infrared colours with
radiation field intensities, for standard filter combinations 
of the ISOPHOT and ISOCAM instruments (on board ISO), and of the 
IRAS survey.

In particular, a filter set covering the
range 60 to 170\,$\,{\mu}$m probes intensities in the ISRF 
ranging from those expected for HII regions to those 
expected in the outskirts of disks of normal galaxies.
By contrast, the MIR colour ratios are almost independent 
of the intensity of the ISRF, since they are determined by
the relative abundance of small, impulsively heated grains.



\section{ISOPHOT MAPPING OBSERVATIONS}
\vspace{-0.3cm}
\section*{OF NEARBY SPIRALS}

ISO represented a big improvement compared with IRAS in 
angular resolution, spectral grasp and sensitivity. Thus, in terms of imaging, 
 the ISOPHOT instrument was able to make a better distinction between diffuse 
and discrete sources and their colours. Another point was the 
extension in wavelength coverage to 200\,${\mu}$m. In fact observations 
of nearby galaxies displayed a wide range of 
FIR colours between various morphological components, implying  
UV/optical/ near-infrared (NIR) interstellar radiation fields with a wide 
range of intensities and/or colours: 
\begin{itemize}

\item {\bf Star formation regions}: 
HII regions (40\,$\le\,T_{\rm D}\,\le\,$60\,K)
and Parent Clouds (15\,$\le\,T_{\rm D}\,\le\,$20\,K)
\item {\bf Nuclear emission} ($T_{\rm D}\,\sim\,$30\,K)
\item {\bf Diffuse Emission}: 
spiral arms and disk (12\,$\le\,T_{\rm D}\,\le\,$20\,K)
\end{itemize}

These features appear to be general to nearby spirals mapped by ISOPHOT. They 
are perhaps best illustrated by the ISOPHOT map of
\object{M~31} (Haas et al. 1998), reproduced in Fig.~2. 
A ring of 10\,kpc radius and a diffuse disk component are clearly seen on this
image at 170\,${\mu}$m. The diffuse  emission can be traced out to a radius of 
22\,kpc, so the galaxy has a similar
overall size in the FIR as seen in the optical bands. 
In addition, there is a faint nuclear source,
which is seen more prominently in HIRES IRAS 60\,${\mu}$m maps at similar 
resolution and in H$_{\alpha}$.
The overall SED, which is also constrained by a 240\,${\mu}$m 
COBE/DIRBE point, can be well described as a  
a superposition of two modified (m=2) Planck curves, with 16 and 45\,K.
The cold dust component at 16\,K arises from both the ring structure 
(30$\%$) and the diffuse disk (70$\%$), illustrating the 
importance of the diffuse emission at least for this example.
The 45\,K component matches up well with HII region complexes in the
ring structure. Associated with each HII region complex are
also compact, cold emission sources (see Fig.~3 of Schmidtobreick, 
Haas \& Lemke 2000) with dust temperatures in the 15 to 20\,K range.
These could well represent the parent molecular clouds in the
star formation complexes which gave rise to the HII regions.
Detailed examination of the morphology of the ring shows a smooth
component of cold dust emission as well as the 
discrete cold dust sources. 

\begin{figure}[htb]
  \begin{center}
    \epsfig{file=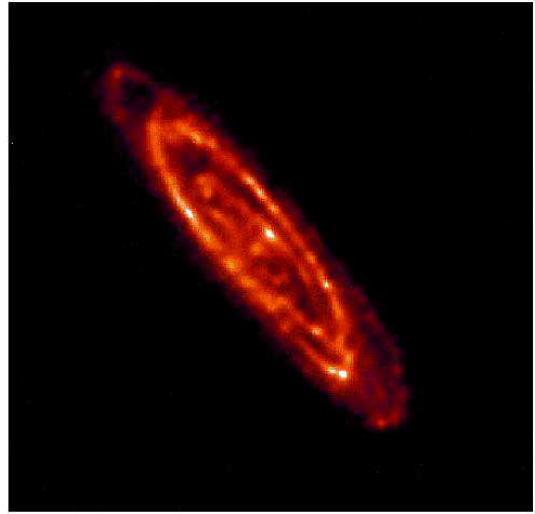, width=7cm}
  \end{center}
\caption{ISOPHOT 170$\,{\mu}$m map of \object{M~31} (Haas et al. 1998), 
with an angular
resolution of 1.3 arcmin. North is towards the top, and East is towards the 
left. The field size is $2.9 \times 2.9$ degrees.
}
\end{figure}

There are also beautiful datasets on normal galaxies which are
more active in star formation than \object{M~31}, 
notably the \object{Small Magellanic Cloud} (type Sm), which was described 
by Bot et al. (2002) and by Wilke et al. (2002), and
\object{M~33} (type Scd; Hippelein et al. 2002).  These galaxies also show a 
prominent diffuse cold
dust component, upon which a warm dust component associated with HII region 
complexes is superimposed. The statistics of the HII region complexes are 
superior to those in \object{M~31}, and, in the case of \object{M~33}, the HII
region complexes exhibit a trend for 
the 60:100 $\mu$m colour temperature to become colder with increasing 
galactocentric distance, as observed in our galaxy.

\begin{figure*}[htb]
\subfigure[]{
\includegraphics[scale=0.6]{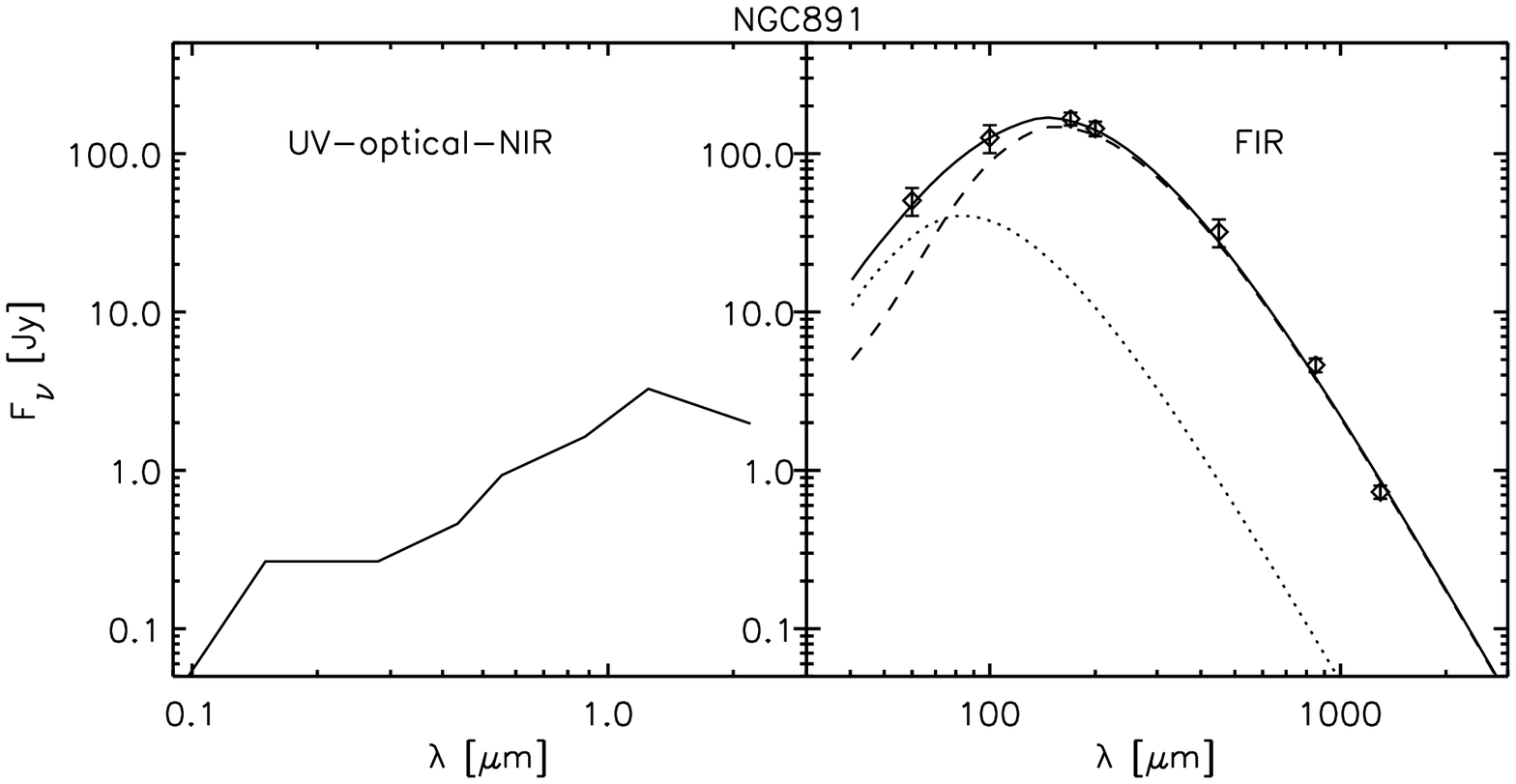}}
\subfigure[]{
\includegraphics[scale=0.49]{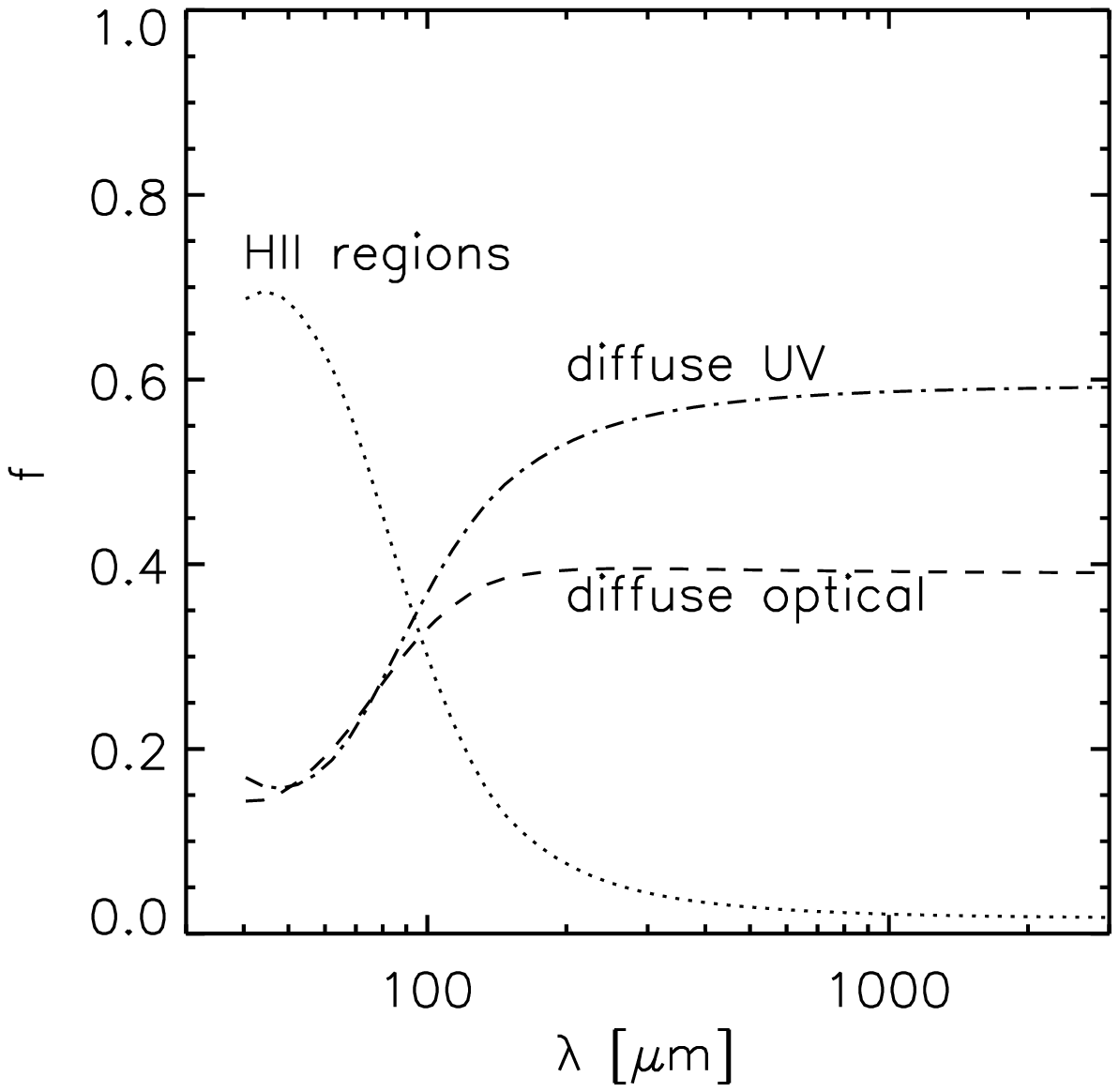}}
\caption{a) The predicted SED of NGC~891 from Popescu \& Tuffs (2002a), 
for a model with $SFR=3.8\,{\rm M}_{\odot}$/yr, $F=0.22$ and 
$M_{\rm dust}=7\times 10^{7}\,{\rm M}_{\odot}$ 
LH panel: the intrinsic emitted 
stellar radiation (as would have been observed in the absence of dust). 
RH panel: the re-radiated dust emission, with
diffuse and HII components plotted as 
dashed and dotted lines, respectively. The data  
(integrated over $\pm 225^{\prime\prime}$ in longitude), are 
from Alton et. al 1998a (at 60, 100, 450 \& 850\,$\mu$m), 
Gu\'elin et al. 1993 (at $1300\,\mu$m) and from Popescu \& Tuffs (2002a) (at
170 \& 200\,$\mu$m).
b) The fractional contribution 
of the three stellar components to the FIR emission, calculated for the case 
of \object{NGC~891}.}
\end{figure*}

Preliminary results from ISOPHOT maps of nearby spiral galaxies were obtained 
by Tuffs et al. (1996) for \object{NGC~6946}, by Hippelein et al. (1996a,b) for
\object{M~51} and \object{M~101} and for a sample of nearby galaxies by Alton
et al. (1998b) and Davies et al. (1999). Claims by the latter two references
that the FIR disks are more extended than the optical disks are not supported 
by the higher linear
resolution maps of M~31 and M~33 nor by the deep map of NGC~891 (Popescu et
al. 2001). However, the maps presented by Alton et al. (1998b) and Davies et 
al. (1999) were not corrected for transient effects of the detector, and 
therefore both the calibration and the derived scale lengths are uncertain.

In general there is a wealth of imaging data in the ISO archive 
still to be exploited, particularly from observations using the
dedicated mapping ``P32'' astronomical observation template.
Excellent photometric images are now becoming available for this mode, 
reduced using new
software techniques (see for example the map of \object{M~101} in 
Tuffs \& Gabriel 2002a,b). 

\section{INTERPRETATION OF FIR EMISSION}
\vspace{-0.3cm}
\section*{FROM NEARBY SPIRALS}

The ISOPHOT imaging studies have directly demonstrated that spiral galaxies
are inhomogeneous. Variations in FIR colours between the observed structures
indicate ISRFs which vary by orders of magnitude in intensity. In general,
this can be attributed to an inhomogeneous distribution of both emitters
and absorbers. Large variations in the colour of the ambient UV-optical
radiation field are also expected, especially if structures are optically 
thick to all or part of the stellar light. Since the absorption
probability of stellar photons by grains is a strong function of wavelength, 
this will also give rise to strong variations in the intensity and colour 
of the FIR emission. A quantitative interpretation of the FIR emission
therefore requires realistic models for the propagation of stellar photons
in galaxian disks, to calculate both the colour and intensity of 
radiation fields illuminating grains.

George Helou described the semi-empirical model of
Dale \& Helou (2002), which assumes a power law distribution of
dust masses over UV-optical radiation energy densities (all of a common 
intrinsic colour). This model can reproduce observed
trends in the MIR-FIR-submm colours of statistical samples, and has also
been used to extract bolometric IR emission and dust 
masses from the ISO data. However, a quantitative interpretation of 
dust emission in terms of star formation rates and star formation 
histories requires a combined analysis of the UV-optical/FIR-submm
SEDs, embracing a self-consistent model for the propagation of the photons.

There are several such models in use which incorporate various
geometries for the stellar populations and dust. We will concentrate 
here on the model of Popescu et al. (2000a), since this is the only model 
where the geometry of the dust and stellar populations
is directly constrained by the optical images. It
is furthermore the only model which has been used to make direct 
predictions for the spatial distribution of the FIR emission for comparison
with the ISO images. Full details of the model are given by Popescu 
et al. (2000a), 
Misiriotis et al. (2001) and Popescu \& Tuffs (2002a). 
In brief, the model
includes solving the radiative-transfer problem for a realistic distribution of
absorbers and emitters, considering realistic models for dust, taking into
account the grain-size distribution and stochastic heating of small grains and
the contribution of HII regions.  The FIR-submm SED is fully determined by just
three parameters: the star-formation rate $SFR$, the dust mass $M_{\rm dust}$
associated with the young stellar population, and a factor $F$, defined as the
fraction of non-ionising UV photons which are locally absorbed in HII regions
around the massive stars.

Here we illustrate this model with the example of the edge-on
galaxy \object{NGC~891}. The model can however be applied to galaxies of any
inclination. The best fit
obtained for the FIR-submm SED of NGC~891 is shown in the right hand panel
of Fig.~3a.


If $L_0$ is the intrinsic UV luminosity
corresponding to unity $SFR$, the relation between the 
FIR luminosity $L_{\rm FIR}^{\rm tot}$ and $SFR$ can be described by:
\begin{eqnarray}
L_{\rm FIR}^{\rm tot}=L_{\rm FIR}^{\rm HII}+
L_{\rm FIR}^{\rm UV}+L_{\rm FIR}^{\rm opt} \nonumber
\end{eqnarray}

\begin{eqnarray}
L_{\rm FIR}^{\rm tot}=SFR \times L_0 \times F + 
SFR \times L_0 \times (1-F) \times G_{uv}+ \nonumber \\ 
+L_{\rm FIR}^{\rm opt} \nonumber
\end{eqnarray}
where $G_{uv}$, which also depends on $M_{\rm dust}$, is the probability 
that a non-ionising UV photon
released into the diffuse dust will be absorbed there.

\begin{figure*}[htb]
\subfigure[]{
\includegraphics[scale=0.5]{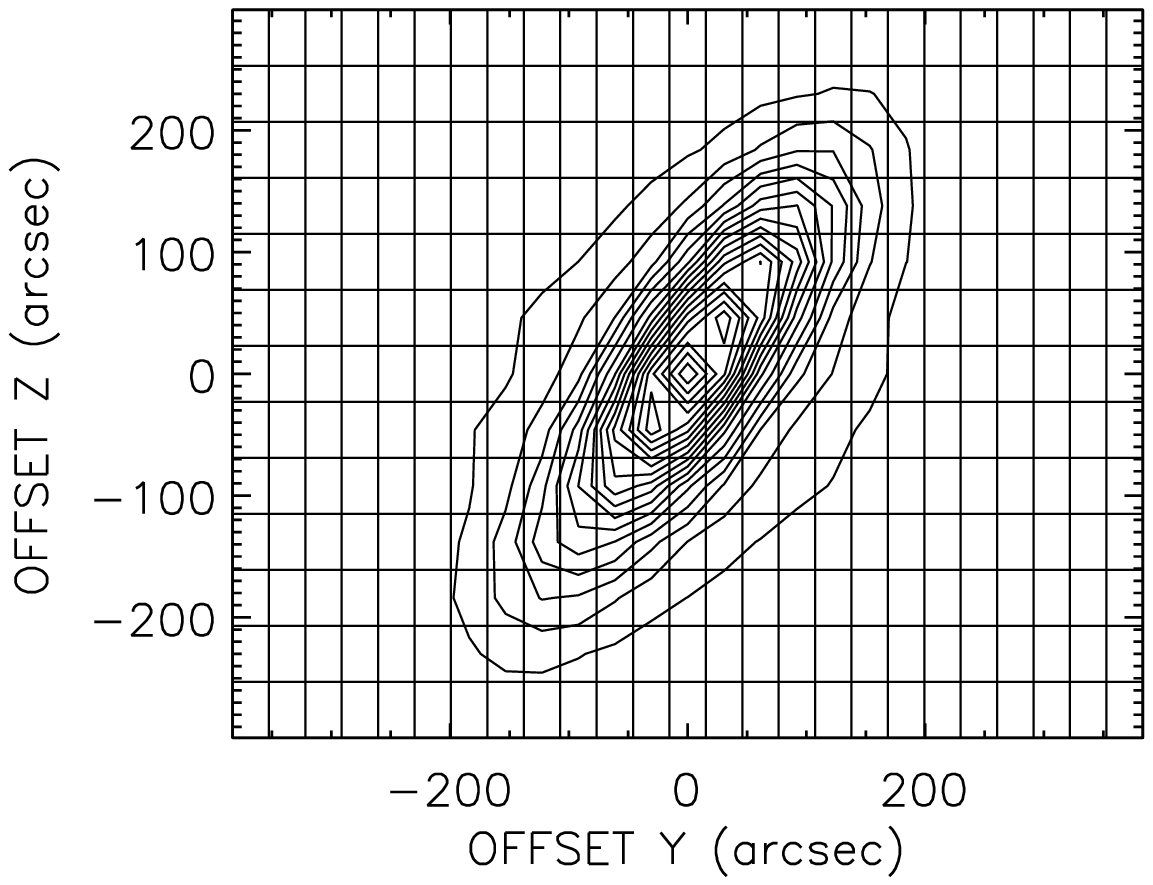}}
\subfigure[]{
\includegraphics[scale=0.5]{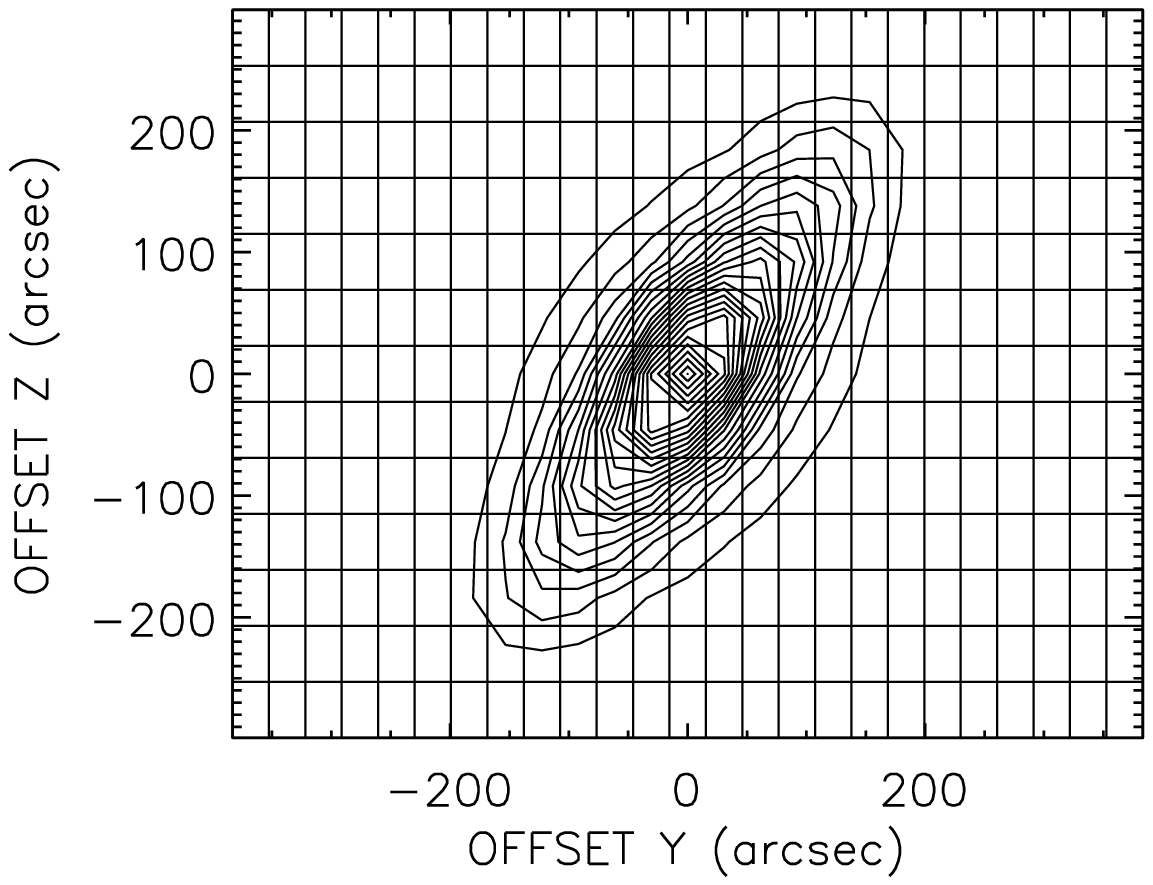}}
\subfigure[]{
\includegraphics[scale=0.5]{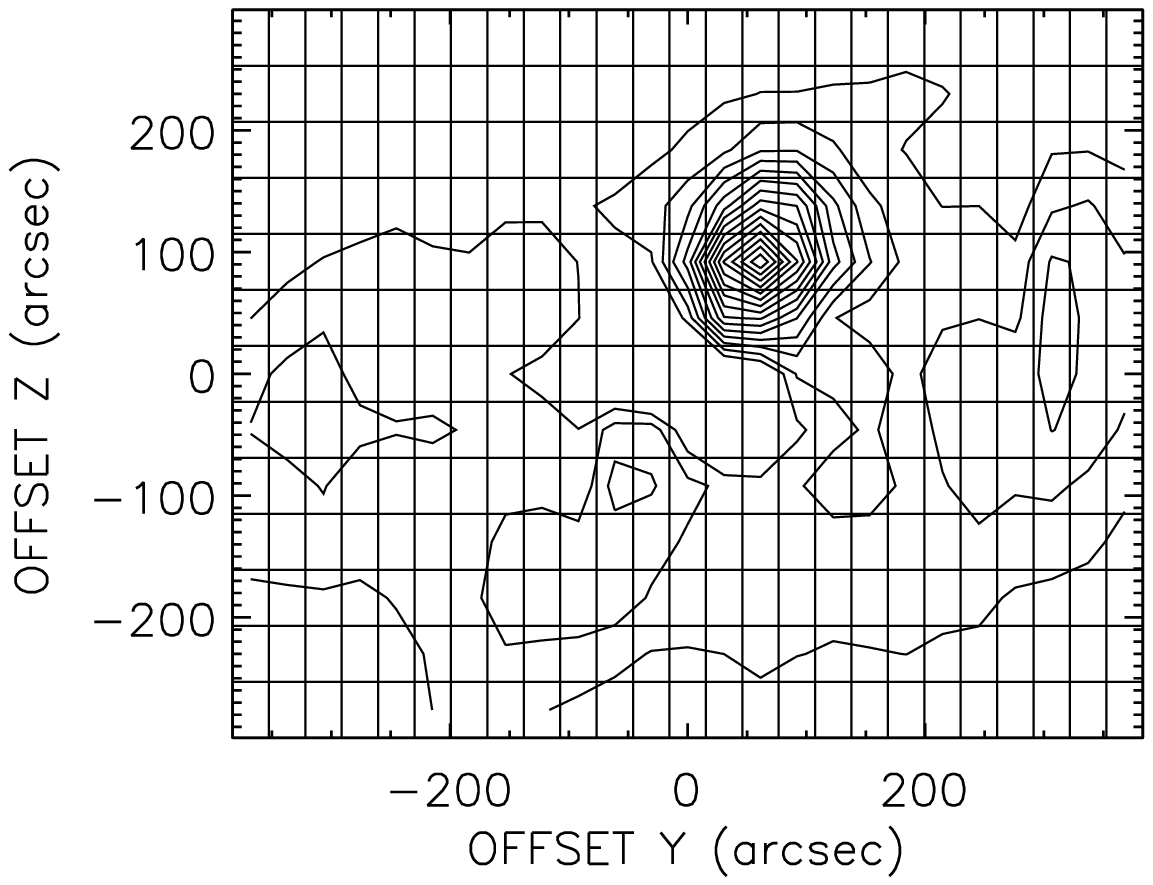}}
\caption{ 
a) Contour plot of the observed brightness distribution of \object{NGC~891} at
170\,${\mu}$m (Popescu \& Tuffs 2002a). The contours are plotted from 10.7 to 
230.0\,MJy/sr in steps 
of 12.2\,MJy/sr. The grid indicates the actual measured sky positions 
sampled at $31 \times 46^{\prime\prime}$ in the spacecraft coordinates 
Y and Z. 
b) Contour plot of the simulated diffuse brightness distribution of 
\object{NGC~891} at 170\,${\mu}$m. The contours are plotted from 10.4 to 
239.7\,MJy/sr in steps of 10.4\,MJy/sr.
c) Contour plot of the observed minus simulated 
diffuse brightness distribution of \object{NGC~891} at 170\,${\mu}$m. 
The contours are plotted from 
2.2 to 46.1\,MJy/sr in steps of 3.1\,MJy/sr. The unresolved source from 
the Northern side of the galaxy and the faint extended source from the 
Southern side account for $8\%$ of the integrated flux density, in
agreement with the model prediction for FIR localised sources at this
wavelength.} 
\end{figure*}

It can be seen that a natural outcome of this modelling technique
is the prediction of a diffuse cold component of dust emission
powered by a combination of non-ionising UV and optical-NIR photons,
and a warm component of dust emission corresponding to the ensemble
of discrete HII regions. This corresponds to what we have seen 
in the ISOPHOT maps of nearby galaxies. 

The model also predicts the 
relative contributions of the young and old stellar populations to
the dust emission as a function of FIR wavelength (Fig.~3b).
An increase in the relative importance of UV heating in the 
diffuse dust emission component towards longer wavelengths 
is apparent. This arises because the coldest grains are those which are in
weaker radiation fields, either in the outer optically thin regions of the disk
or because they are shielded from radiation by optical depth effects. In the
first situation the absorption probabilities
of photons are controlled by the optical properties of the grains, so the UV
photons will dominate the heating. The second situation arises for dust 
associated with the young stellar population, where the UV emissivity far
exceeds the optical emissivity.

This analysis, together with the observational evidence from the
maps produced by ISOPHOT, throws new light on the physical
interpretation of the 60:100$\,{\mu}$m ratio, which is commonly used 
as a gauge of star formation activity (Helou \& Lonsdale 1987). 
This ratio has to be understood in terms of the inhomogeneous nature of disks
in the FIR. Thus, the emission at 
60$\,{\mu}$m mainly traces locally heated warm grains in and around
HII regions, whereas the emission at 100$\,{\mu}$m mainly traces
the emission from the diffuse dust, powered by optical photons
as well as by (non-ionising) UV photons.

A particularly stringent test of the model is to compare its 
predictions for the morphology of the diffuse dust emission component
near the peak in the FIR SED with spatially resolved maps. 
This comparison is shown in Fig.~4a,b, again for the case of \object{NGC~891}.  A remarkable agreement 
between the maps is apparent. 
The residual map between the observed and the simulated maps of the diffuse
component (Fig.~4c) reveals
a compact source on the Northern side of the galaxy (and a faint extended
source seen in the Southern side), with integrated fluxes in
agreement with the model predictions for localised sources.
The excess emission
in the northern side may be a giant molecular cloud complex associated with
one of the spiral arms. The remaining diffuse emission is at a level of 1$\%$
of the peak brightness and is probably attributable to residual detector 
artifacts.

\section{ISOPHOT OBSERVATIONS}
\vspace{-0.3cm}
\section*{OF STATISTICAL SAMPLES}

A total of 31 hours of the ISO mission was dedicated to ISOPHOT pointed 
observations of statistical samples of local universe galaxies. All these 
projects were complementary in terms of selection and observational goals. In
descending order of depth, the published surveys are: the
ISOPHOT Virgo Cluster Deep Survey (63 galaxies), the Coma/A1367 Survey 
(18 galaxies) and the ISO Bright Spiral Galaxies Survey (77 galaxies). In
addition, 115 galaxies have been catalogued from the 
Serendipity Survey. Furthermore, the 
statistical sample of 60 bright galaxies studied spectroscopically by LWS by 
Malhotra et al. (2001) was also observed by ISOPHOT at 170\,${\mu}$m with 
short stares, but the photometry is not yet published.

{\bf The ISOPHOT Virgo Cluster Deep Survey}
(Tuffs et al. 2002; Popescu et al. 2002)
represents the {\it deepest survey} (both in luminosity and surface 
brightness terms) of normal galaxies yet measured in the FIR. A complete
volume- and luminosity sample of 63 gas-rich \object{Virgo Cluster} galaxies 
selected from the Virgo Cluster Catalog \newline (Binggeli, Sandage 
\& Tammann 1985;
see also Binggeli, Popescu \& Tammann 1993) with Hubble types later than
S0 and brighter than $B_{\rm T} \le 16.8$ were measured with ISOPHOT at 60, 
100 and 170\,${\mu}$m. Deep oversampled P32 strip maps covered the entire
optical extent of  each target (down to the 25.5\,mag\,arcsec$^{-2}$ B-band
isophote) as well as adjacent background directions. The faintest detected 
emissions from the 
survey are 50, 40 and 80\,mJy at 60, 100 and 170\,${\mu}$m, respectively. 
The total on target time (ISOPHOT) for the
survey was 20 hours. The sample was also mapped by ISOCAM (Boselli et al. 
1997) and was (in part) observed by LWS (Leech et al. 1999).

The ISOPHOT Virgo Cluster Deep Survey is providing the basis for 
statistical investigations of the FIR spectral energy distributions of gas 
rich galaxies in the local universe spanning a broad range in star-formation 
activity and morphological types, including dwarf systems
and galaxies with rather quiescent star formation activity.

{\bf The Coma/A1367 Survey} (Contursi et al. 2000) consists of 6 spiral and 12 
irregular galaxies having
IRAS detections at 60$\,{\mu}$m. The galaxies were selected to be located 
within 2 or 1 degrees of the X-ray centres of 
\object{Coma} and \object{A1367} clusters, respectively, with emphasis 
on peculiar optical morphologies. Each galaxy was observed in a single
pointing, by chopping between the galaxy and a background direction. The data 
were
taken at 60, 80, 100, 120, 170 and 200$\,{\mu}$m (but only reduced for
120, 170 and 200$\,{\mu}$m) for a total on target time (ISOPHOT) 
of 2\,hr. The sample, also observed with ISOCAM, provided a database of
integrated flux densities for a pure cluster sample of high luminosity spiral 
and irregular galaxies. 

{\bf The ISO Bright Spiral Galaxies Survey} (Bendo et al. 2002a) consists of 
77 spiral and S0 galaxies chosen from the Revised Shapley-Ames Catalog (RSA), 
with $B_{\rm T}\,\le\,12.0$. Almost all are IRAS sources. Mainly an ISOCAM 
survey, the project also used 7\,hr of on target ISOPHOT time to take 
60, 100 and 170$\,{\mu}$m short stares towards the nucleus of the galaxies 
and towards background fields. The sample provides a database of FIR surface 
brightnesses of the central regions of bright spiral galaxies, including S0s. 

The ISO Bright Spiral Galaxies Survey and the ISO-PHOT Virgo Cluster 
Deep Survey represent the principle investigations of optical selected samples
of normal galaxies. It should be emphasised that the main difference between 
them is primarily one of shallow versus deep, rather than field versus cluster,
since by design the Virgo Sample predominantly consists of infalling 
galaxies from the field, and no cluster specific effects could be found.

{\bf The Serendipity Survey} (Stickel et al. 2000) has so far catalogued
115 galaxies with 
$S_{\nu}\,\ge\,2$\,Jy at 170$\,{\mu}$m and with morphological types 
predominantly S0/a\,-\,Scd. This sample provides a database of integrated 
170$\,{\mu}$m flux densities for relatively high luminosity spiral galaxies, 
all detected by IRAS at 60 \& 100$\,{\mu}$m. A more detailed description of
this survey was given in the presentation of Stickel.

\section{INTERPRETATION OF FIR EMISSION}
\vspace{-0.3cm}
\section*{FROM STATISTICAL SAMPLES}

The emerging result of all ISOPHOT statistical studies was 
that the SEDs of normal galaxies in the $40-200\,{\mu}$m spectral range
 require both warm and cold dust emission components to be fitted. 
First indications of this result were given by the multi-filter ISOPHOT 
photometry obtained for 8 inactive spiral galaxies 
(Kr\"ugel et al. 1998; Siebenmorgen, Kr\"ugel \& Chini 1999).
Stickel et al. (2000) and Contursi et al. (2001) found a cold dust 
component for most
of the galaxies in their samples (The Serendipity and the Coma/A1367 samples,
respectively), for which upper limits for the cold dust temperatures were
inferred. The cold
dust component is most prominent in the most ``quiescent'' galaxies, like those
contained in the ISOPHOT Virgo Cluster Deep sample, where the cold dust 
temperatures were found to be broadly distributed, with a median of 18\,K
(Popescu et al. 2002), some $8-10$\,K lower than would have been predicted 
by IRAS. The corresponding dust masses were correspondingly found to be
increased by factors of typically $2-10$ (Stickel et al. 2000) for the 
Serendipity Sample and by factors $6-13$ (Popescu et al. 2002) for the 
Virgo Sample, with respect with previous IRAS determinations. As a consequence,
the derived gas-to-dust ratios are much closer to the canonical value of$\sim
160$ for the Milky Way (Stickel et al. 2000, Contursi et al. 2001), but with a 
broad distribution of values (Popescu et al. 2002).
The FIR properties of the analysed galaxies do not seem to be affected by the
environment (Contursi et al. 2001). 

\begin{figure*}[htb]
\plotfiddle{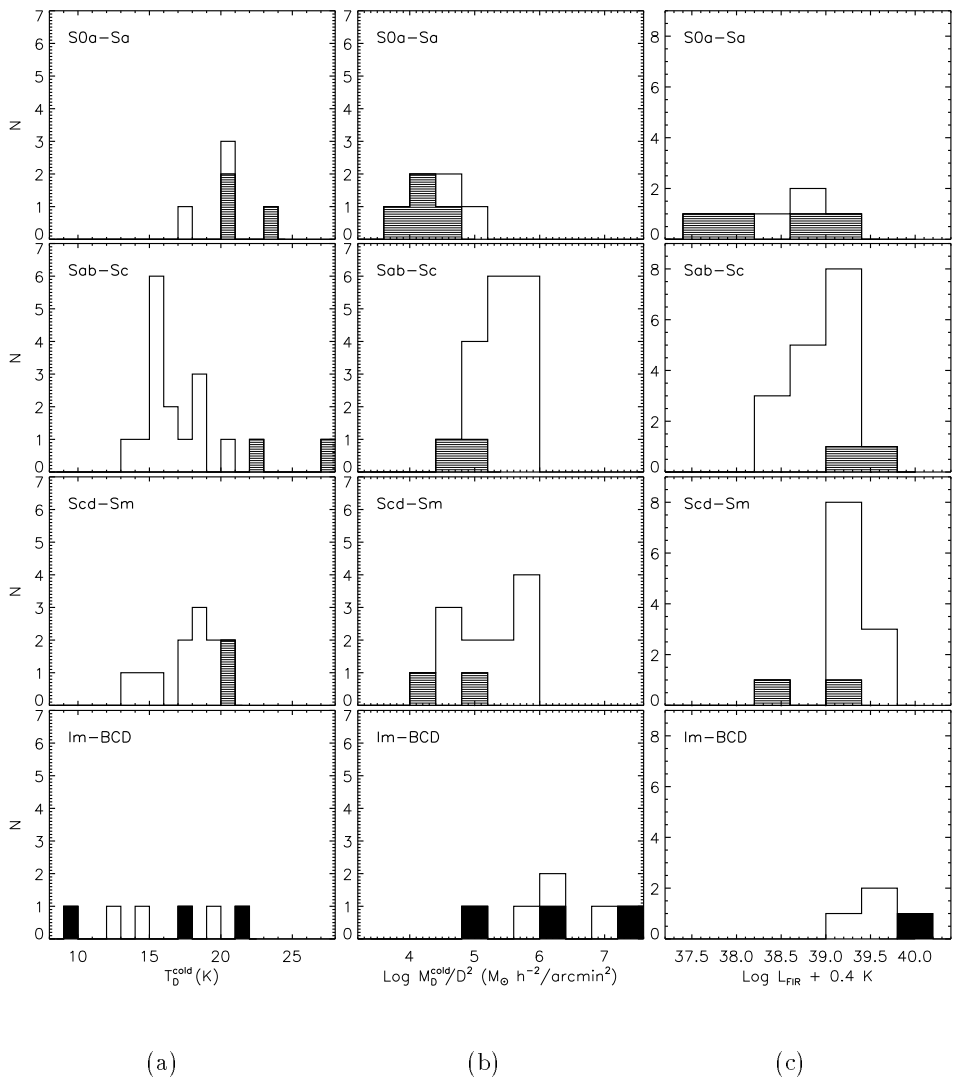}{5.5in}{0.}{95.}{95.}{-240.}{-300.}
\caption{Trends with Hubble type for the ISOPHOT Virgo Cluster Deep sample 
(Popescu et al. 2002). The 
distribution of a) cold dust temperatures 
$T_{\rm D}^{\rm cold}$;
b) cold dust mass surface densities $M_{\rm D}^{\rm cold}/D^2$;
c) normalised FIR luminosity (to the K$^{\prime}$ band magnitude) for 
different Hubble types. The hatched histograms represent the distributions 
for the galaxies with SEDs fitted by only one dust component. The filled 
histograms represent the distributions for the galaxies with detections only 
at 100 and 170\,${\mu}$m.}
\end{figure*}

Of particular interest are the results concerning the trends with Hubble type
found by Popescu et al. (2002) for the ISOPHOT Virgo Cluster Deep 
sample. A tendency was found for the temperatures of the cold dust component
to become colder, and for the cold dust surface densities 
(normalised to optical area) to increase with increasing lateness in the Hubble
type (Figs.~5a,b). A particularly surprising result was the low dust 
temperatures (ranging down to less than 10\,K) and large dust masses 
associated 
with the Virgo Im and Blue Compact Dwarf (BCD) galaxies. Another important 
trend is the increase of
the normalised (to K$^{\prime}$ band magnitude) FIR luminosity as we progress
 from the early to the later Hubble types (Fig.~5c). This result was
later confirmed by Bendo et al. (2002b) for the RSA sample. A related result
was also obtained by Pierini et al. (1999) for the LWS data on 
Virgo galaxies, where a
strong correlation of normalised [CII] emission with H$\alpha$
equivalent widths was interpreted as a trend of increasing star-formation rate
along the Hubble sequence. 

Finally, Popescu \& Tuffs (2002b) also found an
increase of the ratio of the dust emission to the total stellar emitted output
along the Hubble sequence. This correlation is quite strong, ranging from 
typical values of $\sim 15\%$ for early spirals to up to $\sim 50\%$ for some 
late spirals (Popescu \& Tuffs 2002b). This, together with the trend for a
decrease in the temperature of the cold dust, would suggest a trend
of increasing opacities with increasing star-formation activity. 
The extreme BCDs can have even higher 
percentages of their bolometric output re-radiated in the thermal infrared. 
This correlation can be also interpreted as a 
sequence from normal to dwarf gas rich galaxies, with the dwarfs having an 
increased contribution of the FIR output to the total bolometric output. These
findings could be important for our perception of the distant Universe, where,
according to the hierarchical galaxy formation scenarios, gas rich dwarf 
galaxies should prevail. We would then expect these galaxies to make a higher 
contribution to the total FIR output in the early Universe 
than previously expected. 

\section{DUST OUTSIDE GALAXIES}

In some systems dominated by cold dust emission there is evidence that the 
cold dust emission is external to any optically emitting region and/or that 
dust is supplied from an external dust reservoir. 

Thus, the unexpected result that large amounts of cold dust exist in some
Virgo BCDs (Popescu et al. 2002) was interpreted as being indicative of dust 
surrounding the optical galaxy, originating in an external dust reservoir. In 
fact, in two cases direct evidence was found of resolved emission at 170 
micron on scales of up to 10 kpc. The BCD galaxies were found to 
have the highest dust mass surface densities (normalised to
optical area) and the coldest dust
temperatures of the galaxies in the sample.
This is a particularly unexpected result,
since the IRAS observations of BCDs could be accounted for
in terms of dust heated locally in HII regions,
with temperatures of 30\,K or more. To qualitatively account
for the FIR and optical extinction characteristics of BCDs, 
Popescu et al. (2002) proposed two scenarios invoking
collisionally or photon-heated emission from grains originating
in the surrounding intergalactic medium. 
In the one scenario, grains are swept up from a surrounding protogalactic
cloud and heated collisionally in an optically thin wind bubble blown
from the BCD. In the other, the grains are taken to be photon-heated
in an optically thick disk surrounding the optical galaxy. The disk is
indicative of a  massive gas/dust accreting phase which makes dwarf
galaxies sporadically bright optical-UV sources when viewed out of
the equatorial plane of the disk. In both scenarios the dust does not have a
galactic origin, but needs to exist in the immediate vicinity of the galaxies,
where it can either be heated by winds or can accrete into the dwarfs.

Dust outside galaxies has also been discovered
in a deep ISOPHOT survey of a field centred on the giant
elliptical galaxy M86 in the Virgo cluster by Stickel et al. (2002b).
One of the sources of FIR emission seen in the periphery of
the field is extremely cold and has no obvious optical
counterpart. It could trace a dust-rich ``relic'' of the
interstellar media of two spiral galaxies removed in an
interaction as postulated by V\"olk \& Xu (1994). Such objects
could themselves undergo localised episodes of star formation
and could conceivably account for the large dust masses
associated with the Virgo BCDs.

Moving more in the direction of the intragroup medium, the classical 
example
is \object{Stephan's Quintet} (SQ), mapped by ISOPHOT 
using its
oversampling mapping mode P32. The 60\,${\mu}$m map (Sulentic et al. 2001) and
the 100\,${\mu}$m map (Xu \& Tuffs 2002) show the probable detection of
 FIR diffuse emission from the intragroup medium. In particular the
100\,${\mu}$m map shows clear evidence for extended FIR emission in the 
periphery of SQ, its morphology having a striking resemblance to the 
morphology of the diffuse
syncrotron radio emission. Since the diffuse radio emission very probably 
traces a large scale shock in the intragroup medium, the diffuse
 FIR emission is likely to be also associated with the passage
 of the shock front. Such an association was predicted by
 Popescu et al. (2000b) for the case of large-scale accretion shocks
 in clusters.


\begin{acknowledgements}

We thank Dr. J\"org Fischera for providing us with Fig.~1, and 
Prof. Heinrich V\"olk, for informative discussions.

\end{acknowledgements}


\begin{thebibliography}{}

\bibitem[]{} Alton P.B., Bianchi S., Rand R.J., et al., 1998a, ApJ, 507, L125
\bibitem[]{} Alton, P.B., Trewhella, M., Davies, J.I., et al. 1998b, 
A\&A, 335, 807
\bibitem[]{} Bendo, G.J., Joseph, R.D., Wells, M. et al. 2002a, AJ 123, 3067
\bibitem[]{} Bendo, G.J., Joseph, R.D., Wells, M. et al. 2002b, AJ in press
\bibitem[]{} Binggeli, B., Popescu, C.C. \& Tammann, G.A. 1993, A\&AS, 
98, 275
\bibitem[]{} Binggeli, B., Sandage, A. \& Tammann, G.A. 1985, AJ, 90, 1681
\bibitem[]{} Bruzual, G. \& Charlot, S. 1993, ApJ, 405, 538
\bibitem[]{} Boselli, A., Lequeux, J., Contursi, A. et al.
1997, A\&A, 324, L13 
\bibitem[]{} Bot, C. et al. 2002, this volume
\bibitem[]{} Contursi, A., Boselli, A., Gavazzi, G. et al. 2001, A\&A 365, 11
\bibitem[]{} Dale, D. \& Helou, G. 2002, ApJ in press
\bibitem[]{} Davies, J.I., Alton, P., Trewhella, M., Evans, R., \& 
Bianchi, S. 
1999, MNRAS, 304, 495
\bibitem[]{} Gu\'elin, M., Zylka, R., Mezger, P.G., et al., 1993, A\&A, 279, 
L37
\bibitem[]{} Haas, M., Lemke, D., Stickel, M., et al. 1998, A\&A, 338, L33
\bibitem[]{} Helou, G. 2002, this volume
\bibitem[]{} Helou, G. \& Lonsdale, C. 1987, ApJ 314. 513
\bibitem[]{} Hippelein, H., Lemke, D., Tuffs, R.J., et al. 1996a,
A\&A, 315, L79
\bibitem[]{} Hippelein, H., Lemke, D., Haas, M., et al.
A\&A, 315, L82
\bibitem[]{} Hippelein, H. et al. 2002, in preparation
\bibitem[]{} Kylafis, N.D., \& Bahcall, J.N. 1987, ApJ, 317, 637
\bibitem[]{} Kr\"ugel, E., Siebenmorgen, R., Zota, V., \& Chini, R. 1998, 
A\&A, 331, L9 
\bibitem[]{} Laor A., \& Draine B. T. 1993, ApJ 402, 441
\bibitem[]{} Leech, K.J., V\"olk, H.J., Heinrichsen, I., et al.
1999, MNRAS, 310, 317
\bibitem[]{} Lemke, D., Klaas, U., Abolins, J. et al. 1996, A\&A, 315, L64
\bibitem[]{} Malhotra, S., Kaufman, M.J., Hollenbach, D., et al. 2001, ApJ 
561, 766
\bibitem[]{} Mezger P.G., Mathis J.S., Panagia N., 1982, A\&A 105, 372
\bibitem[]{} Misiriotis A., Popescu, C.C., Tuffs, R.J., \& Kylafis, N.D. 
2000, A\&A, 372, 775  
\bibitem[]{} Pierini, D., Leech, K.J., Tuffs, R.J., \& V\"olk, H.J. 1999,
MNRAS 303, L29
\bibitem[]{} Popescu, C.C., Misiriotis A., Kylafis, N.D., Tuffs, R.J., 
\& Fischera, J., 2000a, A\&A, 362, 138
\bibitem[]{} Popescu, C.C., Tuffs, R.J., Fischera, J. \& V\"olk, H.J. 
2000b, A\&A, 354, 480
\bibitem[]{} Popescu, C.C.,  Madore, B.F., Tuffs, R.J., \& Kylafis, N.D. 
2001, AAS198, 76.01
\bibitem[]{} Popescu, C.C. \& Tuffs, R.J. 2002a, Reviews in Modern Astronomy,
vol 15.
\bibitem[]{} Popescu, C.C. \& Tuffs, R.J. 2002b, MNRAS Letters, in press
\bibitem[]{} Popescu, C.C., Tuffs, R.J., V\"olk, H.J., Pierini, D.
\& Madore, B.F., 2002, ApJ 567, 221
\bibitem[]{} Schmidtobreick, L., Haas, M., \& Lemke, D. 2000, A\&A 363, 917
\bibitem[]{} Siebenmorgen, R., Kr\"ugel, E., \& Chini, R. 1999, A\&A, 351, 
495
\bibitem[]{} Stickel, M., Lemke, D., Klaas, U. et al. 2000, A\&A 359, 865
\bibitem[]{} Stickel, M. et al. 2002a, this volume
\bibitem[]{} Stickel, M. et al. 2002b, in preparation
\bibitem[]{} Sulentic, J.W., Rosaldo, M., Dultzin-Hacyan, D., et al.
AJ 122, 2993, 2001
\bibitem[]{} Tuffs, R.J. \& Gabriel, C., 2002a, in Proc. ``ISOPHOT Workshop on 
P32 Oversampled Mapping'', ESA SP-482
\bibitem[]{} Tuffs, R.J. \& Gabriel, C., 2002b, this volume
\bibitem[]{} Tuffs, R.J., Popescu, C.C., Pierini, D., et al. 2002, ApJS 139, 
37
\bibitem[]{} Tuffs, R.J., Lemke, D., Xu, C., et al. 1996,  
A\&A, 315, L149
\bibitem[]{} V\"olk, H.J. \& Xu, C. 1994 Infrared Phys. Technol., 35, 527
\bibitem[]{} Wilke, K. et al. 2002, this volume
\bibitem[]{} Xu, C. \& Tuffs, R.J. 2002, in: ``Proc. ISOPHOT Workshop on P32 
Oversampled Mapping'', ESA SP-482


\end{thebibliography}
\end{document}